# Configurational Electronic States in Layered Transition Metal Dichalcogenides

Jaka Vodeb[1,2], Viktor V. Kabanov[1], Yaroslav A. Gerasimenko[3], Rok Venturini[1,2], Jan Ravnik[1,2], Marion A. van Midden[4], Erik Zupanic[4], Petra Sutar[1] and Dragan Mihailovic[1,2,3]

[1]*Department of Complex Matter, Jožef Stefan Institute, Jamova 39, 1000 Ljubljana, Slovenia*

[2]*Department of Physics, Faculty for Mathematics and Physics, Jadranska 19, University of Ljubljana, 1000 Ljubljana, Slovenia*

[3]*CENN Nanocenter, Jamova 39, 1000 Ljubljana, Slovenia*

[4]*Department of Condensed Matter Physics, Jožef Stefan Institute, Jamova 39, 1000 Ljubljana*

**Mesoscopic irregularly ordered and even amorphous self-assembled electronic structures were recently reported in two-dimensional metallic dichalcogenides (TMDs), created and manipulated with short light pulses or by charge injection. Apart from promising new all-electronic memory devices, such states are of great fundamental importance, since such aperiodic states cannot be described in terms of conventional charge-density-wave (CDW) physics. In this paper we address the problem of metastable mesoscopic configurational charge ordering in TMDs with a sparsely filled charged lattice gas model in which electrons are subject only to screened Coulomb repulsion. The model correctly predicts *commensurate* CDW states corresponding to different TMDs at magic filling fractions $f_m = 1/3, 1/4, 1/9, 1/13, 1/16$. Doping away from $f_m$ results either in multiple near-degenerate configurational states, or an amorphous state at the correct density observed by scanning tunneling microscopy. Quantum fluctuations between degenerate states predict a quantum *charge* liquid at low temperatures, revealing a new generalized viewpoint on both regular, irregular and amorphous charge ordering in transition metal dichalcogenides.**

## Introduction

Recently, scanning tunnelling measurements in transition metal dichalcogenides have revealed irregular charge ordered textures - nanoscale mosaics of charge-ordered domains separated by domain walls which could be created and manipulated by light or by charge injection[1–7].

TMDs form a large family of materials have been of significant interest in the past because they exhibit a variety of charge density wave phenomena (CDWs) at low temperatures[8–10]. Upon doping or the application of pressure, they give rise to a generic phase diagram that usually includes a superconducting state[11] (Fig. 1)[12–21].

The most commonly considered mechanisms for the formation of conventional *long-range ordered CDW states* in TMDs are related to a Fermi surface nesting (FSN) instability[22], electron-phonon coupling[23], and most recently exciton condensation[10,24,25]. In this context it is intriguing that the low-temperature CCDW states in very diverse systems are very similar, yet their proposed formation mechanisms are completely different.

The FSN nesting mechanism for CDW formation is often questioned[26–30]. Eiter et. al.[31] suggested that an enhanced EPC can contribute to or even entirely determine the selection of the ordering vector in $ErTe_3$ for example.

Johannes and Mazin have pointed out that FSN also fails in $NbSe_2$ and $TaSe_2$, and have proposed that EPC can account for the CDW[26]. Similarly, Zhu et. al. have shown that the CDW in $NbSe_2$ can be explained by EPC, predicting the correct CDW wave vector[32].

Coupling of the electrons to the collective mode in the Lee-Rice-Anderson model[33] (LRA) predicts an incommensurate phase and metallic behavior in 1D, but doping by impurities, or interchain coupling predicts the appearance of a gap in the electronic spectrum, which is usually not born out experimentally. Thus, the anomalous peak in resistivity in $1T-TaS_2$ with doping at 8.5 % with Ti and Hf in $1T-TaS_2$[34] could not be understood in terms of the LRA model, and already >1 % of cation doping introduces metallicity - opposite of what is predicted. The absence of observed CDW sliding[35] and recent transport measurements which suggest that topological defects govern transport at low fields[36], and anomalous behavior at higher fields[5] highlights the need to re-address not only the transport problem, but the nature of the state itself.

Tossatti and Fazekas have pointed out the importance of Coulomb interaction in the commensurate (CCDW) state of $1T-TaS_2$, which was proposed to lead to a Mott insulator[37], explaining the large gap and onset of metallicity with doping and pressure[3,4,11,34,38]. On the other hand, band structure calculations have reproduced many of the features of the CCDW state such as the in-plane gap without strong Coulomb correlations[39,40], but fail to reproduce the highly insulating out-of-plane resistivity[41]. In trying to reconcile the viewpoints, Rossnagel suggested that the degree of correlations determines whether the state is an ICCDW or a CCDW in 2D dichalcogenides[10]. More recently, arguments have also been presented for a significant role of the EPC in 1T-TaS$_2$[42,43]. Thus, in spite of the fact that the problem is known since the 1970s, there are many open issues that have not been resolved even in 1T-TaS$_2$, let alone universally in the TMD materials.

Recent scanning tunneling microscopy (STM) experiments have shown that doping introduces densely packed domains and sharp domain walls (DWs) that are not easily understood using the established CDW physics discussed above. Such textures also seem to appear under various external perturbations such as photodoping[5,38,44], charge injection[2–5,45], non-isovalent[46] and isovalent[47,48] transition metal substitution, chemical doping by chalcogen ion substitution[49] or even intercalation of transition metal ions[50]. Apart from DW textures, a few examples of amorphous states have been reported as a consequence of photodoping or charge injection[51] as well as alkali or N$_2$H$_4$ vapor deposition[52,53] and possibly Nb intercalation[47].

The states created through non-equilibrium methods are of particular interest for ultrafast memory devices[5,54–61], but the mesoscopic metastability associated with these phenomena are not easily explained with conventional CDW theory.

Superconducting 2H-NbSe$_2$ exhibits a domain wall structure even in its pristine state with no lock-in transition to a CCDW as in other 2H compounds[62]. Superconductivity thus seemingly coincides with the appearance of a textured state not only in 2H-NbSe$_2$, but also in 1T-TiSe$_2$[15] (and 1T-TaS$_2$[11], another generic feature which conventional theories do not address.

Thus, in spite of intense current interest, it is clear that none of the above-mentioned models which discuss $k$-space interactions in a homogeneous periodic lattice can describe the intricate irregular topologically non-trivial mesoscopic features that universally appear in doped and metastable photoinduced or charge-injected structures[2–4,38,44]. Such discommensurations were discussed theoretically to some extent by McMillan[23] based on Landau theory in 2H-TaSe$_2$, and Nakanishi and Shiba[63,64] for 1T-TaS$_2$, but only as regular structures.

In this paper we approach the problem of describing irregular charge configurational states from the viewpoint of strong correlations between electrons involved in CDW formation in the spirit of the Mott state introduced by Fazekas and Tossatti[37,65,66].

Our approach has its origins in CLG modelling of strongly correlated polarons on a square lattice as applied originally to oxides[67–70] and extended by Brazovskii[71] to investigate the CCDW state of 1T-TaS$_2$. DW formation resulting from doping was suggested by Karpov and Brazovskii[45,71]. Here we examine the phase diagram of a sparse charged lattice gas (CLG) of electrons on a triangular lattice at different filling (i.e. doping) levels using extensive Monte Carlo CLG simulations.

The experimental evidence for local lattice distortions associated with the CDW implies that the electrons are effectively coupled to the lattice, which is usually discussed in terms of polarons, justified in part by the large difference between static and high-frequency dielectric constant that implies substantial screening (as discussed later).

The charged lattice gas (CLG) interpretation of a physical system is typically a very crude oversimplification, yet its intuitive nature is very appealing. In our case it consists of a system of charged repulsively interacting electrons whose motion is slowed down by interaction with the lattice. Despite its simplicity, it has been utilized successfully as a model for a variety of seemingly unrelated physical systems. Superconducting vortices either in Josephson junction arrays[72,73] or superconducting layers in the presence of periodic pinning[74–78] have been studied extensively by Monte Carlo simulations and analytical approaches[79–89] as well as molecular dynamics simulations[90,91]. Adsorbed atoms on a crystalline surface[92–97], colloidal systems on templates[98], organic conductors[99,100], as well as magnetic bubble arrays[101] also exhibit CLG characteristics. Recently, the configuration of a system of polarons in 1T-TaS$_2$ at very low temperatures has been investigated using the CLG model. Remarkably, this model describes well not only the equilibrium state[71] but also two different photoinduced states[45,51]. This success may be considered as justification for attempting to generalize the CLG model for understanding the phase diagram and properties of polaron gases.

Remarkably, we find that the low temperature CCDW states in all the aforementioned systems correspond to electronic crystals, where the filling of the system is one of the many possible *magic fillings* $f_m$. We find that there exists an infinite number of polaron arrangements at non-magic fillings, where the energy gap between the analytically predicted ground state and the first excited long range ordered state can be made arbitrarily small in the thermodynamic limit. Therefore, there exists a whole spectrum of energetically low lying states and this is where an amorphous state is very close in energy to the analytically predicted ground state. Monte Carlo simulations extrapolated into the thermodynamic limit show for a finite amount of cases that this is indeed the case. Moreover, we prove that a triangular lattice cannot exist at $f \neq f_m$ due to the frustration. We conclude that a translationary invariant state with hexagonal symmetry at $f \neq f_m$ indeed needs to exhibit a specific domain wall structure. This is consistent with previously developed theories concerning domain walls[102,103].

From these findings we can draw parallels with a quantum spin liquid. On a triangular lattice a spin liquid is typically understood as a superposition of an infinite number of valence bond states, which all break hexagonal symmetry and have the same energy. The spin liquid ground state then entangles all the states preserving hexagonal symmetry while lowering the energy at the same time[104]. We find a similar situation with the polaron charge configurations in our calculations in the sense that the low lying energy configurations entangle into a *quantum charge liquid*. The MC simulations also confirm a clear absence of a first order phase transition in a wide temperature range in all cases where $f \neq f_m$, which supports the idea that there are two quantum critical points present at $T = 0$ around each magic filling. This is shown on the phase diagram of the CLG on a triangular lattice presented in this paper.

Our conjecture also has important consequences regarding the current paradigm of glass formation[105]. The paradigm states that upon cooling a system down to low temperature, a glass is formed due to the system getting stuck in a local minimum of free energy. There is an overwhelming probability in favor of this phenomenon due to an extensive number of local minima present in the energy landscape of the system. One of the main assumptions in the glass formation paradigm is that the global minimum of the system is the crystalline state, which is challenged by our calculations.

## The Model

The model we employed in this paper is based on the CLG Hamiltonian

$$\mathcal{H} = \sum_{i<j} V(i,j) n_i n_j, \qquad (1)$$

where $n_i$ is the occupational number of a polaron at site $i$ with values either 0 or 1 and $V(i,j) = \frac{V_0 exp\left(-\frac{r_{ij}}{r_s}\right)}{r_{ij}}$ the Yukawa potential that describes the screening, where $V_0 = e^2 \frac{1}{\epsilon_0} a$ in CGS units, $r_{ij} = |r_i - r_j|$, where $r_i$ is the dimensionless position of the $i$-th polaron and $r_s$ is the dimensionless screening radius. The value 1 for both $|r_i|$ and $r_s$ corresponds to one lattice constant $a$ and $\epsilon_0$ is the static dielectric constant of the material. Polarons can occupy only the sites of the underlying triangular lattice with the lattice constant $a$ and the ratio of polarons in the system divided by the number of lattice sites is defined as the filling $f$. We studied the phase diagram of such a system at fixed values of $f$ and $r_s$ as well as the temperature dependence. The idea for the use of polarons in a CLG comes from chapter 4.2 in [106], where a system of repulsive electrons and phonons interacting via the electron-phonon interaction is canonically transformed into a system of interacting small polarons in the strong electron-phonon coupling limit. In this paper we neglect spin effects, assume that the hopping of polarons $\tilde{t} \ll V_0$, justified by the static nature of observed chrges, and assume a screened Coulomb interaction. For further discussion regarding the use of this model see the SI.

## Analytical Considerations

From the structure of the Hamiltonian in Eq. (1) it is obvious that the polarons will always tend to arrange themselves in a triangular lattice with the lattice constant $b = a/\sqrt{f}$, as the most closely packed structure in 2D. However, we need to take into account that polarons can only reside on a triangular underlying atomic lattice (AL). The problem is trivial in the case where the polaron lattice (PL) coincides with the AL. In this case, every point of the PL can be mapped onto a point on the AL.

This condition will be dubbed as the *hexagonality condition* and considered in the following paragraph.

Imagine a 2D plane on which the AL resides. Let the origin of the coordinate system used to describe the points of the plane lie on some AL point and let an arbitrary point on the plane be parametrized by polar coordinates $(r, \theta)$. Also, let an arbitrary AL point be parametrized by polar coordinates $(r_i, \theta_i)$, where $i$ runs through the infinite set of AL points. In order to find all the possible triangular PLs we have to find all the possible triangular lattice unit cells, which map onto the AL. Therefore, we need to find two primitive lattice vectors of a triangular lattice unit cell such that both correspond to some two points on the AL. The hexagonal symmetry of the problem helps because we only need to find one primitive vector. The second one is immediately determined as a $\pi/3$ rotation of the first and will therefore also have to reside on the AL. With the help of the hexagonal symmetry of the problem we can also reduce the search region from the whole plane to only the region of points $\{(r, \theta)|0 \leq \theta < \pi/3\}$. With all of this in mind it becomes apparent that all AL points satisfied by the condition $0 \leq \theta_i < \pi/3$ uniquely determine all the possible triangular lattice unit cell's primitive vectors. With this we have exhausted all of the infinitely many possibilities for a commensurate triangular PL. If we switch to a cartesian coordinate representation for the AL points $(r_i, \theta_i) = x_i(a, 0) + y_i(a/2, a\sqrt{3}/2)$, then the primitive vector of a PL with filling $f$ can be obtained via a simple law of cosines

$$1/f_m = x_i^2 + y_i^2 + x_i y_i. \quad (2)$$

The fillings which satisfy this condition are dubbed as the magic fillings. It is easy to see that $1/f_m = 13$ corresponds to $x_i = 3$ and $y_i = 1$ and $1/f_m = 7$ corresponds to $x_i = 2$ and $y_i = 1$ for example. One interesting fact can be deduced from the mirror symmetry of the problem with respect to the $\theta_i = \pi/6$ axis in the region $0 \leq \theta_i < \pi/3$. Every point $(r_i, \theta_i)$ representing a polaron lattice primitive vector which is not on the $\{(r_i, \theta_i)|\theta_i = 0\}$ or $\{(r_i, \theta_i)|\theta_i = \pi/6\}$ line segments can be reflected over the $\theta_i = \pi/6$ axis in order to obtain a primitive vector $(r_i', \theta_i')$ of a polaron lattice with the same $f_m$ which is distinct from the original. We define them as having two different chiralities. There is no notion of chirality for lattices with primitive vectors on the line segments $\{(r_i, \theta_i)|\theta_i = 0\}$ or $\{(r_i, \theta_i)|\theta_i = \pi/6\}$. In terms of the cartesian coordinate representation the PL has a chirality associated with it whenever the integers $x_i$ and $y_i$ are both non-zero and have different values.

Now consider the case in which the hexagonality condition is not satisfied or in other words, when the system is frustrated. Taking into account that we have just found all the possible cases of hexagonal symmetry in our system, this proves that all the lattice type configurations of polarons must break hexagonal symmetry. None of the 5 possible two-dimensional Bravais lattices possess hexagonal symmetry, except for the triangular lattice. However, by introducing domain walls, it is possible to also preserve the translational and hexagonal symmetry in the system if the domains assume the shape of identical hexagons. This is consistent with previous theoretical work[102,103]. The amorphous states are of course isotropic but there is no long range order present. The two important questions now are how do long range ordered lattice configurations look like and which of them is the most energetically favorable. A lattice must have a periodic unit cell which is how we categorized all the possible configurations. In our analytical considerations we used two lattice energy calculation schemes proposed by Pokrovsky et al.[92] and Arce et al.[95] (see SI). The former was used to evaluate all the possible lattices with only one polaron in the unit cell and the latter with two and three. We stopped at three due the very large number of possible configurations.

## *Results*

### *Monte Carlo Simulations*

We performed MC simulations (for details on the parallel tempering method we employed see SI) with periodic boundary conditions and analytical considerations in order to find the ground state of the system at fixed fillings $f$ ranging from $1/2$ to $1/21$ with the increment of 1 in the denominator. The screening radius $r_s$ ranged from $a$ to $100\ a$ but here we focus on just one value of $r_s = 4.5\ a$ (others are presented in SI). Supplementary Table I (given in the SI) summarizes our results for this value of $r_s$ by listing all the energies per polaron obtained analytically ($E_a$) and compares them with MC simulations ($E_s$). We also list the analytically obtained excitation gap $\Delta$ to the first excited state for each value of $f$. It is clear that all the energies obtained from MC results agree remarkably well with the analytical results up to the accuracy of the MC calculation. This confirms that MC converged very close to the global minimum of free energy. From the order of magnitude of $\Delta$ we can also estimate the order of magnitude of the hopping parameter required to entangle the low lying states $t \gtrsim 0.001\ V_0$ (for an estimation of $\Delta$ and $V_0$ in the case of 1T-TaS$_2$ see SI).

The phase diagram of our model reveals four distinct phases which we define as follows. The crystal phase consists of a perfect long range ordered superlattice of polarons. The domain state is composed of a textured domain structure without long range order and the domains are local formations of the nearest crystal phase. The glass phase exhibits no long range order and is amorphous with glassy dynamics. The liquid phase also does not exhibit long range order, where polarons move around but do not exhibit glassy dynamics. Long range order can be defined via the Fourier transform or pair distribution function (PDF). In the main text we focused on the Fourier transform and the PDF for two cases are discussed in the SI.

Fig. 2 shows a comparison of real space ordering in the low temperature MC simulation and the analytically predicted configurations. In almost all cases where the hexagonality condition (see section Analytical Considerations) is satisfied (at magic fillings $f_m = 1/3, 1/4, 1/7, 1/9, 1/12, 1/13, 1/16, 1/19, 1/21, ...$) we observed a collapse of the system to a long range ordered commensurate triangular lattice. The exceptions are at $f_m = 1/4$ and $1/9$, where topological defects in the form of domain walls prevent full equilibration that is hard to achieve in our MC algorithm (for a discussion of the effect of topological defects on the simulations see SI). Even so, there is no doubt that triangular lattices are still strongly preferred in the two exceptions. In all cases where the hexagonality condition is not satisfied, there is an absence of long range order. This is particularly clearly shown in the Fourier transforms (FT) of the large scale real space (see SI) ordering in Fig. 3.

The orange points represent the FT peaks of intensity of the analytical long range ordered configurations and are superimposed on the FT of the simulated configuration. The FTs confirm the presence of long range order in the magic filling cases. The cases 1/4 and 1/9 stand out due to the presence of intensity also in between the expected peaks, which originates from domain walls. The non-magic filling cases clearly show no long range order due to fact that the peaks of intensity are not located on the expected positions and that they are blurred compared to the sharp analytical peaks. There is also a significant amount of intensity present in between the peaks.

We also performed MC simulations for different values of $r_s$. The overall trend is that the system becomes increasingly harder to equilibrate at larger values of $r_s$. These calculations also reveal the presence of phase coexistence and ordering of domain walls, as well as higher energy amorphous states (see SI for examples).

In Fig. 4 we plot the phase diagram of the CLG model on the basis of MC simulations together with the effective (normalized) gap $\Delta/E_s$. Clearly, the transition temperature increases significantly at magic fillings. This is also where the phase transition is of the first order, and not a glassy transition as it is for $f \neq f_m$. By glassy transition we mean that MC simulations indicate the presence of a weak continuous transition at a certain temperature as is depicted in Supplementary Figures 8 through 11, even though it is obvious from the low temperature configuration snapshots that no ordering is present (see SI). We interpret this transition as the onset of glassy dynamics, where correlations slow down the movement of polarons and eventually freeze the system. Therefore, the transition temperature may be regarded as the interaction scale at which the entropy from all the different low energy long range ordered polaron configurations starts to dominate.

We can speculate that in the quantum case of spinless fermions there are two quantum critical points in the vicinity of each magic filling. Karpov et al.[45] have shown that in the case of $1/f = 13$ the crystal lattice is still present in the system at small negative or positive dopings with the addition of domain walls. If we define the order parameter of the system as the FT peaks corresponding to the 1/13 lattice, then it is obvious that the order parameter does not disappear immediately under doping. It is very natural to assume that this is the case for all magic fillings. However, at the same time our MC simulations clearly show that if we dope the system enough, the first order phase transition becomes a glassy transition (see SI). This implies that the order parameter disappears in this case.

Likely, there is a quantum critical point somewhere in between. This is the case for positive and negative doping but the two critical points are not necessarily symmetric around each $f_m$. Also, it is possible that at low enough fillings the two critical points merge and even overlap and therefore disappear, as the magic fillings get closer and closer to one another.

In order to avoid the scaling of the gap $\Delta$ with $f$ we overlayed the plot of the relative gap $\Delta/E_s$ on top of the phase diagram. The peaks in the plot agree well with the peaks in the phase diagram, which explains the emergence of a first order transition at $f_m$. There is a well pronounced free energy minimum at $f_m$, which is not true in the frustrated case. The cases 1/4, 1/9 and 1/16 stand out due to the fact that a square lattice is energetically very close to the triangular lattice. This is probably true for every non-chiral lattice (see Analytical Considerations) and it is consistent with the emergence of domain walls in simulations for 1/4 and 1/9 fillings as the walls probably cost very little energy compared to the chiral cases. The plot also clearly shows that even in the case of our limited analytical analysis the frustrated cases of 1/6, 1/10 and 1/14 exhibit a gap which is almost zero (of the order of $10^{-8}$). We believe that if our analytical analysis would be extended to polaron lattices with 4 or more polarons per unit cell, a further reduction in the value of the gap will occur in all the other non-magic filling cases.

*Exact Diagonalization of a Charged Lattice Gas With Quantum Tunneling*

In order to substantiate our claim regarding the quantum charge liquid, we carried out exact diagonalization calculations using the Lanczos method in order to find the ground state. The code used in the calculations was from the ALPS initiative, version v2.3b1[107–110]. We were interested in how does the introduction of quantum effects such as the hopping of electrons alter the charge order compared to the classical case. The model we used is the model of spinless fermions with a fixed filling and both nearest and next-nearest neighbor interaction

$$\mathcal{H}_{sf} = -t \sum_{<i,j>}(c_i^\dagger c_j + c_i c_j^\dagger) + \sum_{<i,j>} V_1 n_i n_j + \sum_{<<i,j>>} V_2 n_i n_j \quad ; \quad \sum_i n_i = N, \tag{3}$$

where $c_i$ ($c_i^\dagger$) is the annihilation (creation) fermion operator on the $i$-th triangular lattice site, $n_i = c_i^\dagger c_i$, $t$ is the hopping parameter, $V_1$ the interaction between nearest neighbors, $V_2$ between next-nearest neighbors and $N$ is the number of polarons in the system. We set the system size to $5 \times 5$ and $N = 5$ in order to explore the frustrated case of $f = 1/5$ and also set $V_1 = 10$, $V_2 = 5$ and varied $t$. Supplementary Figure 13 shows the density correlation function $\langle n_i n_j \rangle$ for the classical case $t = 0$ and two quantum cases $t = 0.5$ and $t = 10$.

The correlation function clearly shows that for small values of $t$, the interaction driven charge ordering dominates in the system due to the fact that the correlation functions in the $t = 0$ and $t = 0.5$ cases are practically the same. This is not the case for larger values of $t$, where a metallic phase is present as one would expect for a system where kinetic energy of electrons dominates. These calculations by themselves do not prove the existence of a QCL because system size limitations limit the accuracy. Indeed, the density correlation function of the small 5 × 5 system is substantially different from that from classical Monte Carlo simulations. However, they do confirm that the presence of a large number of nearly degenerate configurations in the frustrated cases most likely plays a significant role.

*Experimental Realizations of Predicted Charge Orders*

We have collected all of the experimentally studied systems for which our model is applicable in Table 1.

| System | $f$ | Phase |
| --- | --- | --- |
| 2H-Fe$_{0.33}$TaS$_2$[111] | 1/3 | crystal |
| 1T-TiSe$_2$ | 1/4 | crystal |
| N$_2$H$_4$/1T-TaS$_2$[112] | 1/4 | crystal |
| N$_2$H$_4$/2H-TaS$_2$[112] | 1/4 | crystal |
| 1T-Cu$_{0.08}$TiSe$_2$[50] | 1/4.2 | domain state |
| 2H-Fe$_{0.33}$TaSe$_2$[111] | 1/4 | crystal |
| 2H-Fe$_{0.33}$NbSe$_2$[111] | 1/4 | crystal |
| Alkali/1T-TaS$_2$[113] | ~ 1/8 | glass |
| N$_2$H$_4$/1T-TaS$_2$[112] | ~ 1/8 | glass |
| 2H-TaS$_2$[114] | 1/9 | crystal |
| 2H-TaSe$_2$[114] | 1/9 | crystal |
| 2H-NbSe$_2$[62] | ~ 1/9 | domain state |
| Cu/1T-TaS$_2$[115] | 1/9 | crystal |
| Alkali/1T-TaS$_2$[113] | 1/9 | crystal |
| N$_2$H$_4$/1T-TaS$_2$[112] | 1/9 | crystal |
| PD 1T-TaS$_2$ | 1/11 | glass |
| 1T-Nb$_{0.1}$TaS$_2$[47] | ~ 1/11 | possible glass |
| PD 1T-TaS$_2$ | 1/12.6 | domain state |
| 1T-TaSeS | 1/12.6 | domain state |
| 1T-Ta$_{0.99}$Fe$_{0.01}$S$_2$ | 1/12.6 | domain state |
| 1T-Ti$_{0.07}$Ta$_{0.93}$Se$_2$[48] | 1/12.6 | domain state |
| 1T-Nb$_{0.04}$TaS$_2$[47] | ~ 1/13 | domain state |
| 1T-Nb$_{0.07}$TaS$_2$[47] | ~ 1/13 | domain state |
| 1T-TaS$_2$ | 1/13 | crystal |
| 4Hb-TaS$_2$[114] | 1/13 | crystal |
| 1T-TaSe$_2$ | 1/13 | crystal |
| 4Hb-TaSe$_2$[114] | 1/13 | crystal |
| 1T-NbSe$_2$[114] | 1/13 | crystal |
| 1T-VSe$_2$[114] | 1/16 | crystal |

**Table 1** A collection of all the experimentally studied systems for which our model is applicable. Every system in left column for which we were able to extract the filling $f$ within the scope of our model is associated with its $f$ and corresponding theoretical phase in the middle and right column, respectively. In the six examples in which this was not possible, we specify the corresponding theoretical phase and only the closest $f_m$ in the case of a domain state and the approximate $f$ in the amorphous case with a ~.

Perfect polaronic crystals according to [111–115] are realized for the following fillings (systems): $f_m = 1/3$ (2H-Fe$_{0.33}$TaS$_2$), 1/4 (1T-TiSe$_2$, 2H-Fe$_{0.33}$TaSe$_2$, 2H-Fe$_{0.33}$NbSe$_2$, N$_2$H$_4$/2H-TaS$_2$, N$_2$H$_4$/1T-TaS$_2$), 1/9 (2H-TaS$_2$, 2H-TaSe$_2$, Alkali/1T-TaS$_2$, Cu/1T-TaS$_2$, N$_2$H$_4$/1T-TaS$_2$), 1/13 (1T-TaS$_2$, 4Hb-TaS$_2$, 1T-TaSe$_2$, 4Hb-TaSe$_2$, 1T-NbSe$_2$) and 1/16 (1T-VSe$_2$).

In Fig. 5 we show a number of very different examples of charge ordering for which we could determine the system's filling and in which doping is achieved by different mechanisms: photodoping, *isovalent* transition metal or chalcogen substitution, *non-isovalent* substitution (i.e. chemical doping) or interstitial doping. Most of the images were obtained using an Omicron 4-probe LT STM at 5 K. The exceptions are the TaSeS image (obtained using a Specs JT-STM at 1 K), Cu intercalated TiSe$_2$ (images taken from [50]) and Ti doped TaSe$_2$ (images taken from [48]). The experimental setup and sample growth were described previously [38].

The first case (Fig. 5a,e,i) of photoexcited 1T-TaS$_2$ shows the pristine state ($f = f_m = 1/13$), the lightly photoexcited "hidden" state[116] and moderately photoexcited "amorphous" state[51]. With light photoexcitation, the domain state is reached, while for moderate photoexcitation, an amorphous state is reached. The measured values of $f$ are 1/13, 1/12.6 and 1/11 respectively. For $f \neq f_m$, the density is determined by counting the polarons in the respective images using standard procedures described in [51]. In the amorphous state, $f$ falls in between two magic numbers ($f_m = 1/9$ and 1/12).

The second case (Fig. 5a,b,h) shows non-isovalent chalcogen substitution from S to Se in 1T-TaS$_x$Se$_{2-x}$, where the main effect was recently suggested to be effective doping through broken-symmetry-induced hybridization upon local buckling[49]. The end cases 1T-TaS$_2$ (Fig. 5a) and 1T-TaSe$_2$ (Fig. 5b) both have $f = f_m = 1/13$, while the intermediate case is superconducting 1T-TaSeS ($x = 1$, Fig. 5h) that clearly exhibits domain walls, as expected for

$f \neq f_m$. The estimated filling (by counting polarons) is $f \approx 12.6$, which is similar to the photoexcited hidden state.

The third example (Fig. 5a,d) shows Fe-doped 1T-Ta$_{1-x}$Fe$_x$S$_2$, which implies chemical doping. The undoped 1T-TaS$_2$ (Fig. 5a) has a filling of $f = 1/13$ and the estimated filling in the doped case is $f \approx 1/12.6$ ($x = 0.01$, Fig. 5d).

The fourth example (Fig. 5b,g) is isovalent TM ion substitution of 1T-Ti$_x$Ta$_{1-x}$Se$_2$[48]. The filling is $f = 1/13$ in the undoped ($x = 0.07$, Fig. 5b) and $f \approx 1/12.6$ in the doped (Fig. 5g) case.

The fifth example presented here (Fig. 5c,f) shows doping of 1T-TiSe$_2$ by Cu intercalation[50]. While undoped 1T-TiSe$_2$ (Fig. 5c) shows 1/4 charge order, with 1T-Cu$_{0.08}$TiSe$_2$, domain walls are clearly seen (Fig. 5f). The estimated density of polarons from the doping level is $f \approx 1/4.2$, which suggests hole doping.

The final example that we highlight here is doping by alkali or N$_2$H$_4$ vapor deposition on the surface of 1T-TaS$_2$[52,53] where a state emerges with $f = 1/8$, which was dubbed as a commensurate $c(2\sqrt{3} \times 4)rect$ CDW phase. It appears as a result of competition of three energetically equivalent 1/8 rectangular lattices, which are rotated by 120 degrees relative to each other. The resulting structure is thus composed of small domains of superlattices, which are rotated relative to one another with domain walls in between. For a visual representation see Supplementary Figure 3.

These examples were chosen because they display diverse doping mechanisms, but show generic behavior with doping as predicted by the phase diagram in Fig. 4.

There are also two other compounds worth mentioning, which could not be presented either in Fig. 5 or in the list of polaronic crystals. The case of intercalated 1T-Nb$_x$TaS$_2$[47] exhibits a 1/13 polaronic crystal at $x = 0$, a domain state at $x = 0.04$ and $x = 0.07$ and a possible amorphous state at $x = 0.1$. However, we could not reliably determine the system's filling and therefore we excluded this case from the analysis in Fig. 5. It should also be noted, that the role of structural disorder is not clear in this case. The second case is 2H-NbSe2, which even in its pristine form does not exhibit a perfectly commensurate CDW. However, the work of Chatterjee et al.[62] has shown that there are indeed local commensurate structures present with phase slips in between. Our polaron picture interprets this as a domain state, which is very intriguing, as it is the only domain state present in the absence of external perturbations to the system.

## *Discussion*

The presented calculations suggest a remarkably simple and generic model for understanding the phase diagram of TMDs with particular focus on correlated configurational ordering. Our viewpoint intentionally neglects other interactions that may lead to competing or intertwined orders, for example the long wavelength (~ 50 nm) chiral states created by laser pulses[1,38]. We neglect the fact that a Fermi surface instability may win at higher temperatures, when thermal fluctuations destroy local textures caused by Coulomb correlations, or the polarons themselves become unbound. A good example is 1T-TaS$_2$, where Fermi surface nesting gives a nesting wavevector that gives a real-space charge modulation which is close to the 1/13 structure, and a series of transitions is observed from high-temperature incommensurate order to low-temperature correlated Mott state with 1/13 filling. Many other TMDs exhibit an incommensurate CDW preceding the CCDW[114], which means that additional interactions, such as FSN need also be considered in such cases to describe the high-temperature orders. The model does not address the electronic structure of the resulting compounds, and the emergent phases are not universally insulating as one would expect in a static polaronic crystal. For example, 2H-TaSe$_2$ is metallic in the CCDW phase down to the lowest temperatures, while others show anomalous resistivity[41].

The electronic band structure and transport properties may be modelled by DFT calculations of the folded structures[40,49], sometimes achieving quite good agreement with experiments such as angle-resolved photoemission for example. Nevertheless, transport properties still remain anomalous and are often not well understood. Resistivity relaxation measurements suggest that quantum configurational reordering processes may play a role in transport[41].

The microscopic effect of doping in all the different compounds is not yet well understood and is not universal. Substitution seems to cause the formation of more polarons in the system. Intercalation also seems to cause the formation of more polarons, except for the 1T-Cu$_{0.08}$TiSe$_2$ case. The role of fluence in photodoping has also not yet been fully explored, however the work of[1,51] suggest that larger fluences lead to more polarons being formed. We also note that the isovalent Se for S substitution in 1T-TaS$_{2-x}$Se$_x$ effectively introduces doping though strain induced by the size mismatch, but the end members (1T-TaS$_2$ and 1T-TaSe$_2$) both have $f_m = 1/13$[17].

Dai et al.[111] show that substantial Fe doping of 2H-TaS$_2$, 2H-TaSe$_2$ and 2H-NbSe$_2$ causes the formation of CDWs with smaller periodicities than the undoped materials (from $3a \times 3a$ to $2a \times 2a$, or $\sqrt{3}a \times \sqrt{3}a$, where $a$ is the lattice constant) above a certain treshold Fe doping level.

The apparent outlier of all the compounds is 1T-TiSe$_2$, where exciton condensation is the proposed mechanism for CCDW formation[25]. However, this approach on its own cannot explain the presence of the observed DW structure which emerges under pressure[15] or doping with transition metal ions[48], nor does it take into account Coulomb interactions. We propose that 1T-TiSe$_2$ might be a good example of cooperation between strong polaron correlations enhanced by exciton crystallisation that favours the CCDW with $f_m = 1/4$.

The model presented in this paper is consistent with the observations in the work of Dai et al.[111] on Fe doped 2H-TaS$_2$, 2H-TaSe$_2$ and 2H-NbSe$_2$. The change of the superlattice type in experiments can be explained as the change of the magic filling of the present polaronic crystal. They also report on distortions in the superlattice, which is consistent with the domain wall state as well as the amorphous state in between magic fillings.

One interesting fact is the absence of the 1/7 superlattice predicted by the model, which has not been observed anywhere else. Doping also seems to increase the importance of the role of correlations in the system due to the onset of commensurability from incommensurability in 2H-TaSe$_2$ and 2H-NbSe$_2$.

From the presented analytical calculations and the MC results it is clear that for all magic filling cases there exists a global minimum of free energy that corresponds to a triangular polaron lattice. It might seem that similar magic number filling might appear also on square lattices[81], or other more complex periodic lattices, but this is not the case: Triangular packing is a global energy minimum because a triangular arrangement of polarons with an isotropic interaction is densest possible packing in 2 dimensions. So a square lattice may give rise to a local minimum of sparse periodic filling, but cannot yield a state that is a global energy minimum. Indeed, as far as we aware, there are no examples of square regular sparse CCDWs on a square (tetragonal) lattices amongst either chalcogenides or oxides. Instead, various frustrated textures form, such as stripes, nematic phases and bipolarons[67,68,81].

The induced frustration in all the non-magic filling cases causes the system to break hexagonal symmetry and an amorphous state becomes very close in energy to the analytical ground state. The resulting hyperuniform jammed packing is the densest possible maximally correlated state in 2D that is not periodic. We interpret the fact that an amorphous state is so close in energy to the analytical ground state to be a consequence of the presence of a large amount of entropy even at very low temperatures. There seem to be a large number of states available to the system at low temperatures and they are mixed into an amorphous state even if the temperature is only slightly above zero. We conjecture that in every case where the hexagonality condition (see section Analytical Considerations) is not satisfied, the gap between the ground state and the first excited state can be made arbitrarily small.

From the fact that there are many broken-symmetry states present at low energies, we can speculate that for the quantum case of the CLG model (spinless fermions), where one introduces even a very small amount of nearest neighbor hopping $t$ ($t \ll V_0$) and fixes the system's filling, the ground state will entangle the low lying energy states ($|\psi_i\rangle$) into a quantum charge liquid ($|\psi\rangle = \sum_i c_i |\psi_i\rangle$, where $\langle\psi|\psi\rangle = 1$) analogue of a quantum spin liquid with no long range order. The system would therefore have an amorphous ground state which breaks the current paradigm of glass formation by which a crystalline state is the eventual ground state of a glassy system. It is important to note that $t$ needs to be small enough so that correlations are still dominant in the system but large enough to retain entanglement between the low-lying states. At larger values of $f \neq f_m$ a so called "pinball liquid" state was proposed to be the ground state[100,117]. However, due to limitations in the system size of quantum calculations, we believe that true long range order is still elusive in the two abovementioned studies. A much larger system size would be necessary in order to achieve conclusive results. An amorphous ground state has already been proposed within the context of frustration between two different length scales of the interaction[118,119]. In our case frustration emerges from the competition between the interaction-driven polaron lattice and the underlying atomic lattice on which the polarons reside (see Analytical Considerations). The QCL in both the domain state or the glassy regime would in the theoretically ideal case exhibit a homogeneous real space structure, assuming quantum coherence across the whole system. However, in actual experiments impurities would force the system to choose a specific configuration as observed in [38,51]. Therefore, only states local to the energy of the chosen state would be susceptible to entanglement. An interesting implication which follows is that any kind of observed polaron tunneling or even tunneling between different configurational states could be explained by the QCL concept.

Finally, let us estimate the macroscopic tunneling rate $\tau$ between the near-degenerate configurational states. The barrier tunnelling rate depends on the barrier height $E$ as $\tau \sim exp(-W\sqrt{E})$, where $W$ is a constant. The energy of the system $E$ is proportional to the number of particles involved $N$, and therefore in the thermodynamic limit as $N \to \infty$, the tunneling rate $\tau \to 0$. However, there may be processes in which a finite number of particles $N_{eff}$ are involved, so $\tau$ is finite. If the configurations do not differ significantly, the tunneling rate may be significant, and observable.

Reconfiguration processes should show a cross-over from tunneling to thermally-activated classical behavior above some threshold temperature. Experimentally this should be observable as a cross-over from temperature-independent tunneling to thermally activated behavior at some characteristic temperature. Such processes may be observed when a small number of polarons are involved in a slight reconfiguration of domains, for example.

In the amorphous state jamming may inhibit the relaxation process[51], and the system is exceptionally stable, in agreement with observations.

## Conclusions

Remarkably, the generic physical picture which emerges appears to predict the CCDWs wavelengths at: $f_m = 1/3$ (2H-Fe$_{0.33}$TaS$_2$), 1/4 (1T-TiSe$_2$, 2H-Fe$_{0.33}$TaSe$_2$, 2H-Fe$_{0.33}$NbSe$_2$, N$_2$H$_4$/2H-TaS$_2$, N$_2$H$_4$/1T-TaS$_2$), 1/9 (2H-TaS$_2$, 2H-TaSe$_2$, Alkali/1T-TaS$_2$, Cu/1T-TaS$_2$, N$_2$H$_4$/1T-TaS$_2$), 1/13 (1T-TaS$_2$, 4Hb-TaS$_2$, 1T-TaSe$_2$, 4Hb-TaSe$_2$, 1T-NbSe$_2$) and 1/16 (1T-VSe$_2$).

To our knowledge, other $f_m$ values are not experimentally known until now, but particularly the $1/7$ state has a sufficiently distinct energy minimum that it might exist.

The fact that doping away from $f_m$ leads to the formation of charged domain walls presents a straightforward interpretation of where doped charges reside. Simply counting sites in STM images (Fig. 5) gives the doping level which agrees with the chemical composition. Of course, one needs to take into account that diverse domain wall patterns are present which are both electron-rich and hole rich[3,38,45].

At half-way points between successive $f_m$ values the amorphous phase appears to be experimentally realized by photoexcitation at $f = 1/11$[51] or possibly Nb intercalation[47]. The phase diagram thus suggests that in the quantum case of the CLG (including quantum hopping) two quantum critical points ($T = 0$ transitions) could emerge, separating each $f_m$ state from the amorphous states in between. Due to the extreme near degeneracy present in the domain wall or the glassy case, we speculate that an entangled quantum charge liquid might also emerge in the quantum case of the CLG.

It appears that the ground state of many TMDs forms crystalline polaron order, which seems to not be correlated with Fermi surface properties at high temperatures. Which $f_m$ appears apparently cannot be easily correlated with the band structure. Fermi surface nesting thus appears to be an additional effect, not the underlying cause for the formation of the low-temperature commensurate state. Instead, within our approach, $f_m$ is determined primarily by an interplay between the magnitude of the electron-phonon interaction that determines the electron bandwidth in relation to the Coulomb repulsion to satisfy the Mott criterion. The beauty of this viewpoint is that it gives insights into the formation of domain walls and the amorphous order that cannot be obtained from conventional reciprocal space models or DFT.

As the charge ordering within the CLG model emerges from the presence of strong screened Coulomb correlations present in the system, all of the listed systems would therefore appear to be dominated by strong correlations at low temperatures. The temperature where the picture breaks down is related to either the point where polarons become mobile or polaron dissociation, which may be roughly estimated from their binding energy[120]

$$E_p = \frac{1}{2\kappa} \int_{BZ} \frac{d^3q}{(2\pi)^3} \frac{4\pi e^2}{q^2}, \qquad (4)$$

where $\kappa^{-1} = \epsilon_\infty^{-1} - \epsilon_0^{-1}$ and $\epsilon_\infty$ and $\epsilon_0$ are the high frequency and the static dielectric constant respectively. Using the published values for the lattice parameters from[121] and the static[36] and high frequency[122] dielectric constant, we obtain $E_p \approx 0.19\ eV$ in the case of 1T-TaS$_2$. The temperature which corresponds to this energy is about $2200\ K$, which is much higher than any reported phase transition temperature in these materials.

We conclude by highlighting the model's predictions in relation to observed peculiarities which are not easily otherwise understood. The most obvious is the ubiquitous and immediate emergence of domain wall structures under various kinds of doping of TMDs. The prediction of an amorphous state, as recently observed under specific and unusual photoexcitation conditions[51] in between magic fillings is another. The model also finally explains the anomalous peak in resistivity with Ti or Hf doping of 1T-TaS$_2$ which occurs at $f_m = 1/12$ ($p \simeq 0.085$) that has been a puzzle since the 1970s[37] and has led to the original proposal of a Mott state[37,65]. Another beautiful experiment that begs for an explanation is alkali-vapour doped 1T-TaS$_2$ by Rossnagel et al.[52,53], where continuous tuning the electronic correlations by surface Rb deposition reveals a 1/8 state. The experiments of Dai et al.[111], that reveal a transition from one magic filling to another under substantial doping show that it is possible to move between magic fractions along the doping axis within a single compound.

Finally, given that superconductivity appears in between $f_m$ and coincides with the appearance of domain textures, this work suggest that new superconductors may be searched for in other areas of the phase diagram that were not previously investigated for superconductivity. This may also reveal some valuable insight in answering the question how the appearance of textures is related to superconductivity - a question that has wider and persistent significance in cuprates and other textured high-temperature superconductors.

## Acknowledgments

We wish to thank Peter Karpov and Tomaz Mertelj for the useful discussions. Funding from ERC-2012-ADG_20120216 "Trajectory" and the Slovenian Research Agency (program P1-0040, program P1-0099, young researcher P0-8333, young researcher P0-07586 and young researcher P0-7589) is acknowledged.

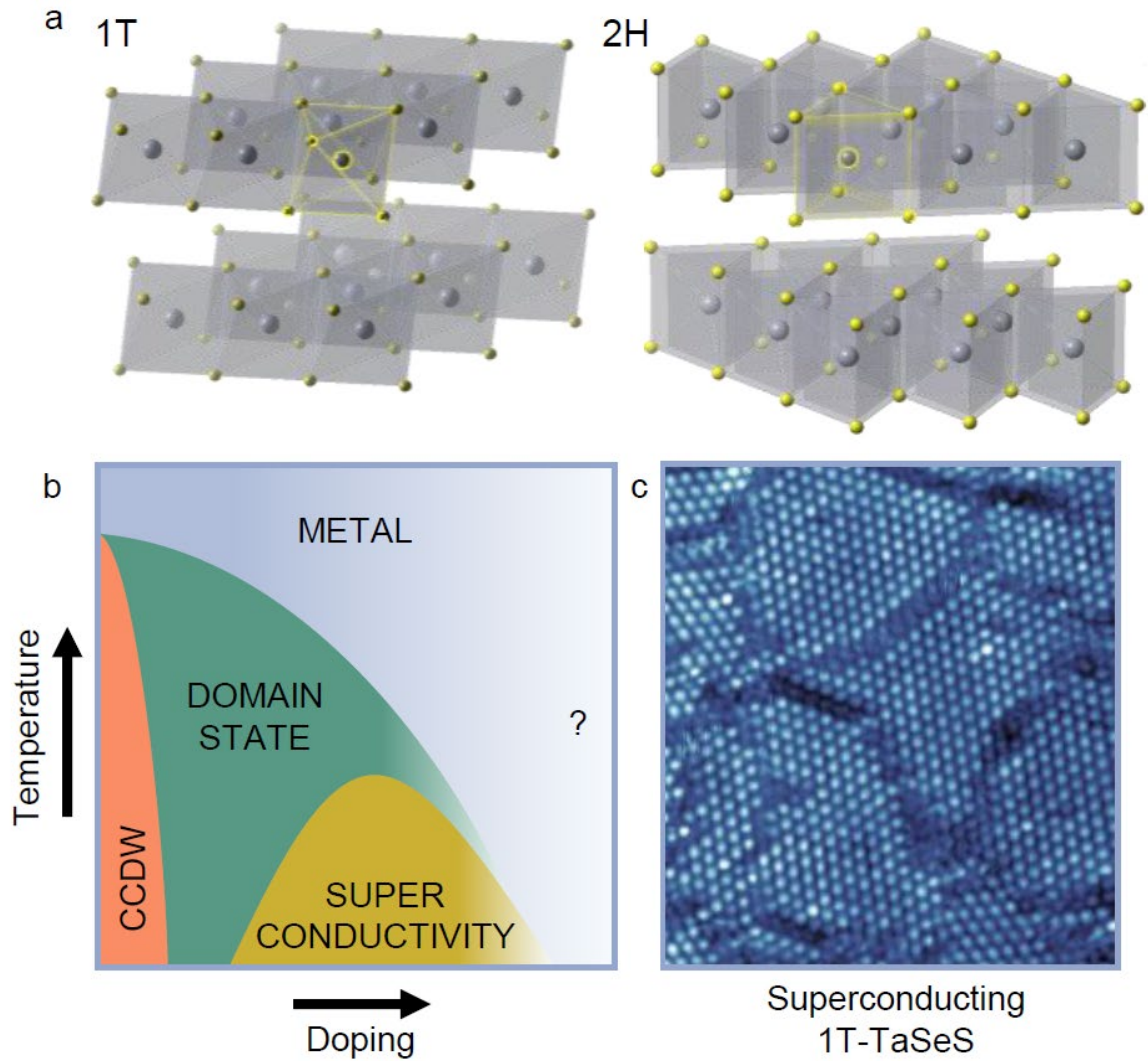

**Fig. 1** A schematic representation of a typical crystal structure in layered materials, which exhibit a CCDW. **a** This specific case shows two polytypes (1T and 2H) of $TaS_2$. The unit cell is emphasized in yellow, the yellow spheres represent sulphur and the gray spheres are metal atoms. **b** The generic phase diagram of doped TMDs. STM experiments show that doping immediately results in the formation of domains which persist in the superconducting state. The ? indicates that this part of the phase diagram is poorly understood. **c** The domain structure measured at $1.1\ K$ in the superconducting state of 1T-TaSeS ($T_c = 3.2\ K$).

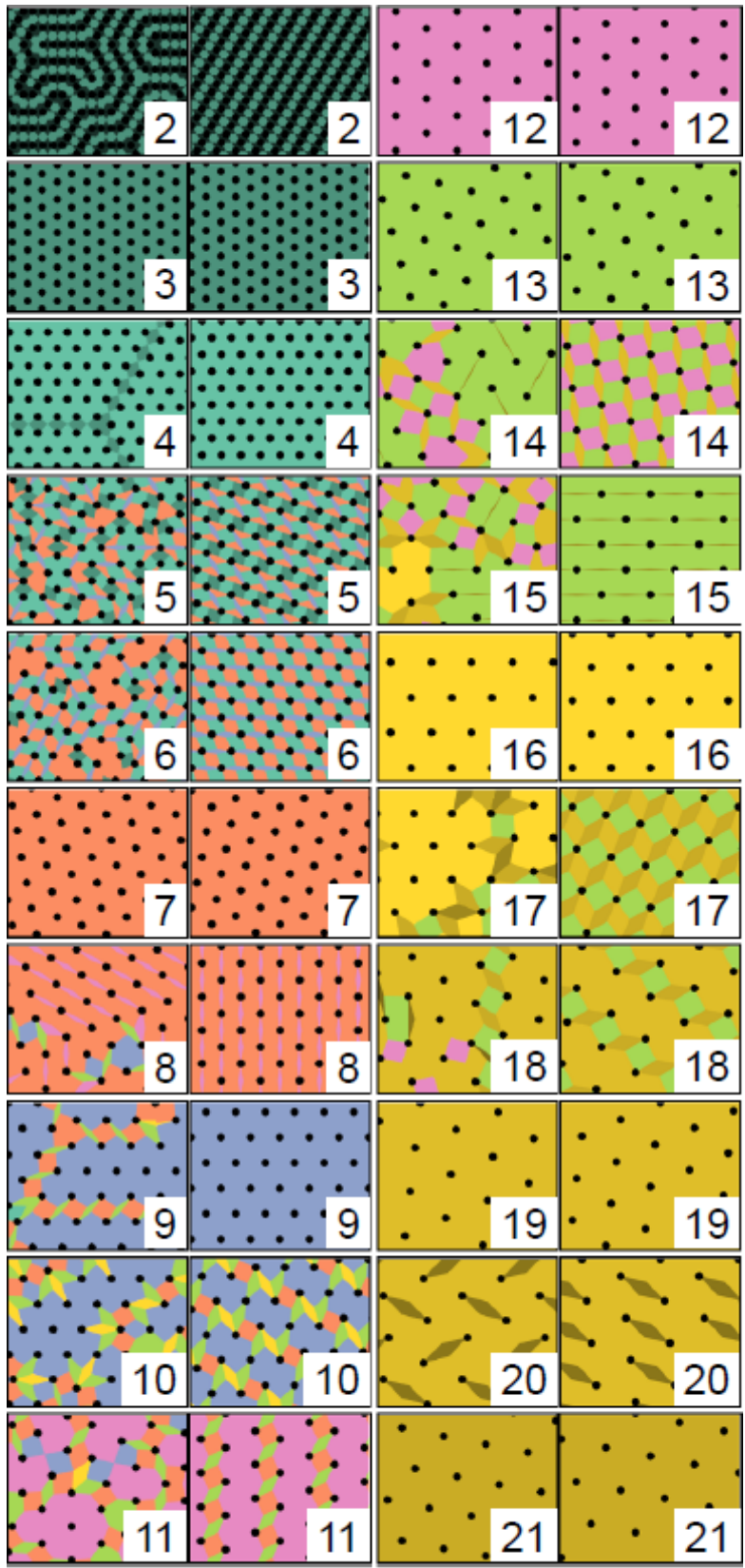

**Fig. 2** Comparison of Monte Carlo simulation snapshots at low temperatures with analytical predictions of ground states at different values of $1/f$, where $f$ is the filling of the system. The figure is organized in two columns of pairs with the same value of $1/f$ denoted on each image. The left image in the pair is always the simulation snapshot and the right is the analytical prediction. The colors in the image represent distances between neighboring polarons and are presented according to the legend at the bottom of the figure. For details regarding the tiling algorithm see SI.

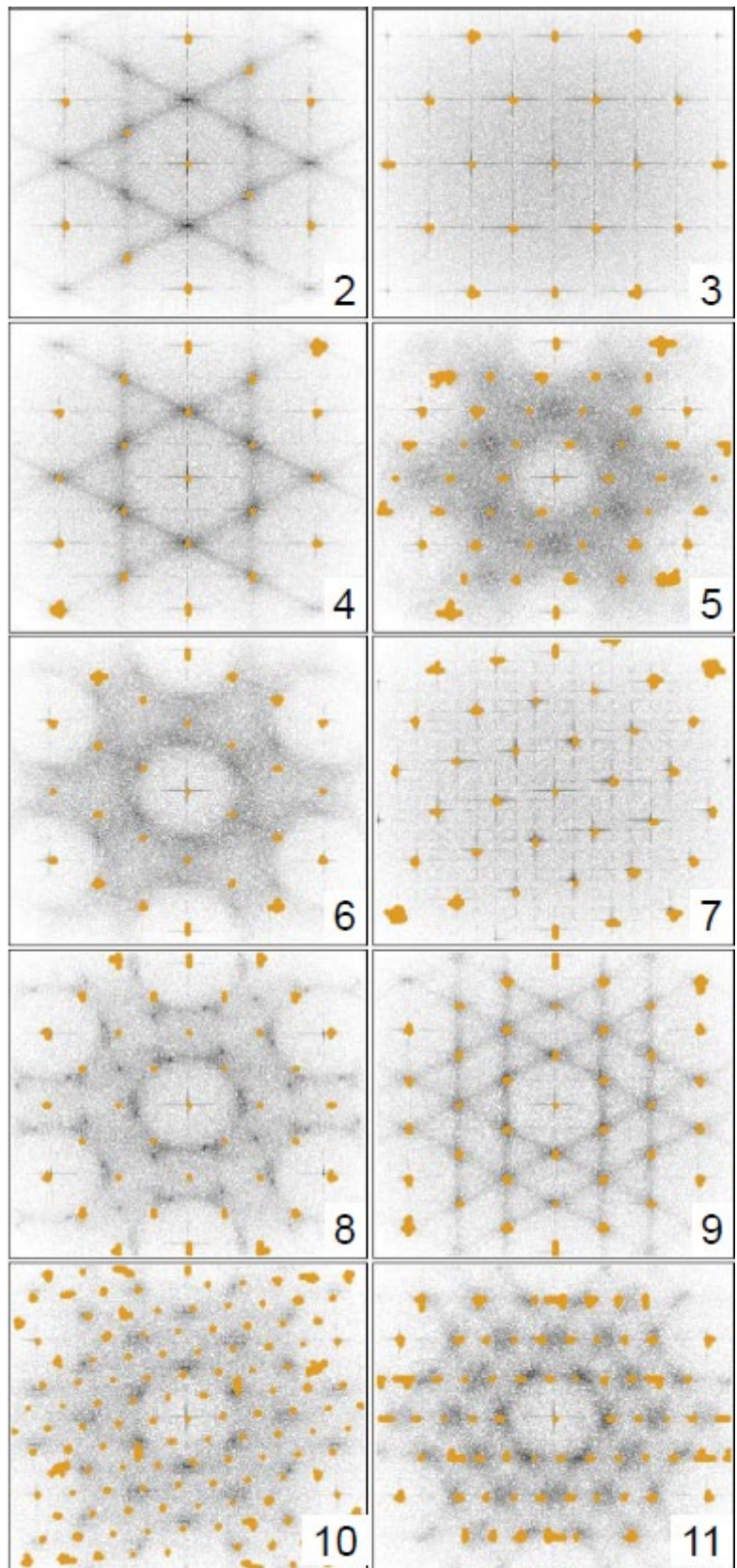

**Fig. 3** Comparison of Fourier transforms of the large scale images in Fig. 2. The peaks of intensity from the analytical cases are overlayed in orange on top of the simulated cases. The number in each image represents the value $1/f$ like in Fig. 2.

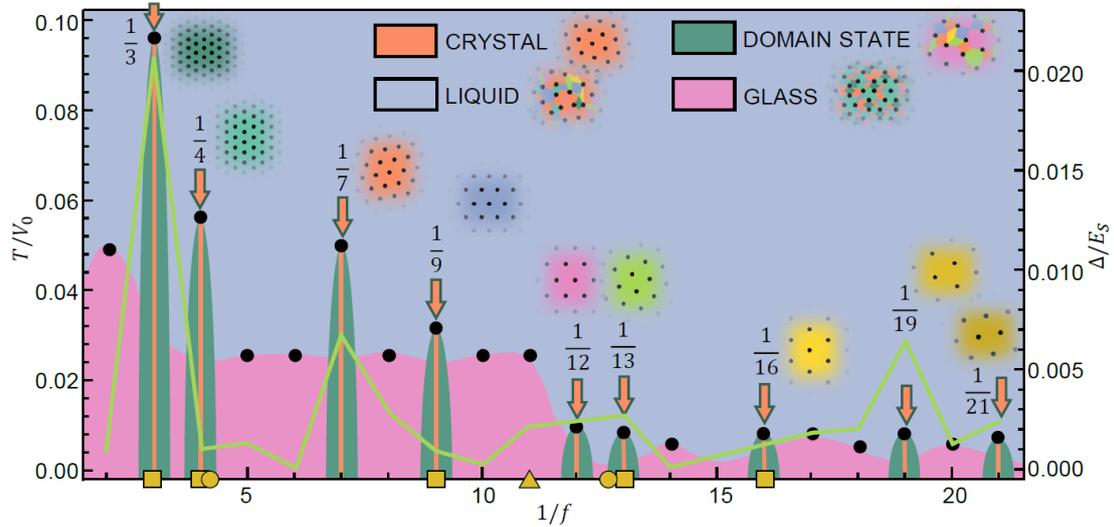

**Fig. 4** Charge Lattice Gas phase diagram of temperature $T$ versus $1/f$. The liquid, crystal, domain and glass phases are colored in blue, orange, dark green and pink respectively and the boundary between them represents the phase transition. Magic fillings are marked with arrows. The green line is the calculated gap between the ground state and the first excited state. The color coding of the commensurate phases is the same as in Fig. 2. The experimental realizations of the polaron crystal phase are marked with yellow squares: $f_m = 1/3$ (2H-Fe$_{0.33}$TaS$_2$), $1/4$ (1T-TiSe$_2$, 2H-Fe$_{0.33}$TaSe$_2$, 2H-Fe$_{0.33}$NbSe$_2$), $1/9$ (2H-TaS$_2$, 2H-TaSe$_2$), $1/13$ (1T-TaS$_2$, 4Hb-TaS$_2$, 1T-TaSe$_2$, 4Hb-TaSe$_2$) and $1/16$ (1T-VSe$_2$), the domain states with yellow circles: $f \approx 1/12.6$ (photodoped 1T-TaS$_2$, 1T-Ti$_{0.07}$Ta$_{0.93}$Se$_2$, 1T-TaSeS, 1T-Ta$_{0.99}$Fe$_{0.01}$S$_2$), $f \approx 1/4.2$ (1T-Cu$_{0.08}$TiSe$_2$) and the glass phase with a yellow triangle: $f \approx 1/11$ (1T-TaS$_2$). They are presented in more detail in figure Fig. 5 and main text.

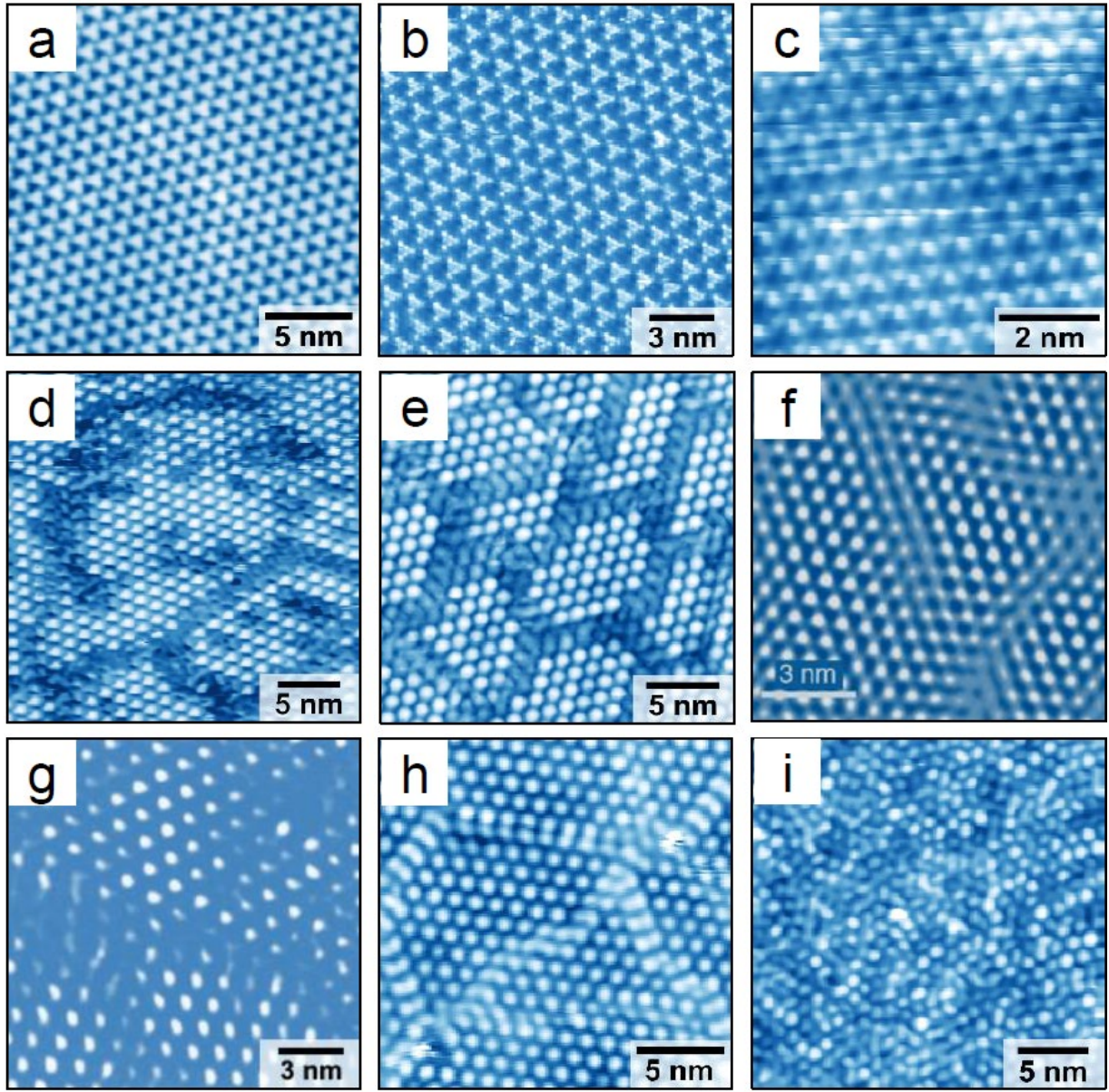

**Fig. 5** Scanning tunneling microscope images for selected compounds showing the effect of doping, arranged in three successive groups of polaron crystals, domain states and a glassy state. The first three cases **a-c** represent polaronic crystals in 1T-TaS$_2$ ($f_m = 1/13$), 1T-TaSe$_2$ ($f_m = 1/13$) and 1T-TiSe$_2$ ($f_m = 1/4$) respectively. The latter cases **d-h** represent domain states of 1T-Ta$_{0.99}$Fe$_{0.01}$S$_2$ ($f \approx 1/12.6$), photodoped 1T-TaS$_2$ (fluence $F \simeq 1\ mJ/cm^2$, $f \approx 1/12.6$), 1T-Cu$_{0.08}$TiSe$_2$ ($f \approx 1/4.2$), 1T-Ti$_{0.07}$Ta$_{0.93}$Se$_2$ ($f \approx 1/12.6$) and 1T-TaSeS ($f \approx 12.6$) respectively. The last case **i** represents a glassy state in photodoped 1T-TaS$_2$ (fluence $F > 3.5\ mJ/cm^2$, $f \approx 1/11$).